\documentclass[superscriptaddress,nofootinbib,floatfix, amsfonts, aps]{revtex4}

\usepackage{amsmath}
\usepackage{bm}
\usepackage{graphicx}
\usepackage{mathrsfs}
\usepackage{footmisc}
\usepackage{multirow}
\usepackage{graphicx}
\usepackage{booktabs, ctable}
\usepackage{xcolor, color, framed}

\newcommand{\erf}{\textrm{erf}}
\newcommand{\dd}{\textrm{d}}

\usepackage[normalem]{ulem}  

\renewcommand\sout{\bgroup \color{red} \uldepth=-.5ex \ulset}
\begin{document}

\title{Fast Heating Dissociation of $\Upsilon(1S)$\ in Heavy Ion Collisions at RHIC}
\author{Yunpeng Liu} 
\author{Baoyi Chen}
\thanks{Email: baoyi.chen@tju.edu.cn}
\affiliation{Department of Applied Physics, School of Science, Tianjin University, Tianjin, 300072, China}

\date{\today}

\begin{abstract}
   With the adiabatic assumption in the cooling process, we discussed a new mechanism on $\Upsilon(1S)$ suppression that is due to the fast heating process at the early stage of the fireball instead of its finite decay width in finite temperature medium  produced in the heavy ion collisions. We calculated the transition probability after the fast heating dissociation as a function of the temperature of the medium and the nuclear modification factor in central collisions, and found that the suppression is not negligible at RHIC, even if the width of $\Upsilon(1S)$\ vanishes.
\end{abstract}

\maketitle

The phase diagram of quark matter is one of the most interesting topics in high energy nuclear physics. At high temperature and/or high baryon density, the state of the quark-gluon plasma (QGP) has been predicted and widely studied both in theory~\cite{Hagedorn:346206, Karsch:2000kv, Ding:2015ona, Pisarski:1983ms, Fukushima:2010bq, Stephanov:1998dy, Kharzeev:2007tn} and in experiments~\cite{Gonin:1996wn,Luo:2017faz, Adcox:2001jp, Adare:2006ti,Chatrchyan:2011sx}. Lots of synchrotrons and colliders are built for heavy-ion collisions to produce the QGP. However, it is difficult to measure the temperature of the fireball directly, especially at its very early stage, because of its small size and short lifetime. From the statistical model~\cite{Andronic:2009qf}, one can extract the temperature from the spectrum or the yield of light hadrons. It seems that strange particles freeze-out earlier than pions~\cite{bellwied2019extracting}, which means that they carries information of the fireball at earlier time. Heavy quarkonia may even survive the QGP due to their large binding energies and carry information from the early stage of the fireball. However, they can hardly be thermalized, either in kinetics or in chemistry. Anomalous suppression of $J/\psi$ was suggested as a signal of the formation of QGP very early~\cite{Matsui:1986dk}, which actually contains different processes such as color screening~\cite{Karsch:2005nk}, gluon scattering~\cite{Peskin:1979va, Xu:1995eb, Chen:2018dqg,Yao:2018sgn},  and quasi-free scattering from quarks~\cite{Grandchamp:2001pf, Yao:2018sgn}. It is found that the inverse process, that is the regeneration of $J/\psi$\ from charm quarks in QGP also plays an important role in relativistic heavy collisions at RHIC and higher colliding energies~\cite{Thews:2000rj, Grandchamp:2001pf, Yan:2006ve, Yao:2018nmy}, which makes the story more complicated.

Although it is difficult to measure the temperature of the fireball at early time from the momentum distribution of heavy quarkonia, the sequential dissociation model~\cite{Karsch:2005nk} offers another way to take heavy quarkonia as thermometers of the fireball, which assumes that a heavy quarkonium survives if and only if the temperature of the fireball is above the dissociation temperature $T_d$ of the quarkonium. Most of other dynamic studies focus on the dissociation rate (or the width) of a heavy quarkonium at a certain temperature $T$~\cite{Thews:2000rj, Grandchamp:2001pf, Yan:2006ve, Zhu:2004nw, Song:2007gm, Zhao:2011cv, Brambilla:1999xf, Escobedo:2013tca, CS:2018mag}. As a matter of fact, the sequential dissociation model can also be regarded as a special case of the dynamic models with a dissociation rate that is infinitely large above $T_d$\ and zero below $T_d$.

However, even if the dissoicaiton rate of a heavy quarkonium vanishes at finite temperature, quarkonium suppression can still happen in heavy-ion collisions because of the fast heating process at the very early stage of the bulk medium, which is the main effect we try to discuss in this paper. In relativistic heavy ion collisions, the fireball reaches its highest temperature within $1$\ fm/c, and cools down for much longer time to freeze-out finally. For simplicity, we treat the initial heating process as a sudden process, and the cooling process as a very slow process. According to the adiabatic theorem, the yield of heavy quarkonia keeps constant during the cooling process if the width of the quarkonia is negligible. Therefore the suppression is mainly due to the transition from initial heavy quarkonia to heavy quarkonia in the hot medium. In this case, we can map the nuclear modification factor of heavy quarkonia to the temperature of the fireball after the fast heating process directly. As discussed in the previous paragraph, $J/\psi$\ at RHIC and at colliders with higher beam energies does not fit such a model, since both the scattering dissociation and the regeneration are very important. Therefore we will consider $\Upsilon(1S)$ instead in the following discussion and calculations, since both its dissociation rate and its regeneration rate in medium are small at RHIC energy.~\cite{Liu:2009wa, Du_2017}

The evolution of the wave function $\psi({\bm r},t)$ of  $\Upsilon(1S)$\ at time $t$ and the relative radius ${\bm r}$ between the bottom quark and the anti-bottom quark  can be described by the Schroedinger equation
\begin{eqnarray}
   i\partial_t \psi({\bm r},t)&=& \left[-\frac{1}{m_b}\nabla^2+V(r,T(t))-i\Gamma({\bm r}, T)\right]\psi({\bm r},t),\label{eq_1}
\end{eqnarray}
where $m_b$\ is the mass of a bottom quark, and we have taken $\hbar=1$. In the above the thermal fluctuation~\cite{Katz:2015qja} is neglected. The corresponding stationary radial Schroedinger equation for a give temperature $T$ writes
\begin{eqnarray}
   \left[-\frac{1}{m_br}\frac{\dd^2}{\dd r^2}r+\frac{l(l+1)}{m_br^2}+V(r,T)-i\Gamma(r,T)\right]\psi_r(r)&=& E\psi_r(r),
\end{eqnarray}
where $\psi_r(r)$\ is the radial wave function of $\Upsilon(1S)$, and $E$\ is the eigen energy of $\Upsilon(1S)$. For $\Upsilon(1S)$, the azimuthal quantum number is $l=0$.
 To focus on the new mechanism, we neglect the particle scattering process and  take the in-medium width $\Gamma=0$. The potential $V$\ is taken in the form of a screened Cornell potential~\cite{Satz:2005hx}
\begin{eqnarray}
   V(r)&=&  -\frac{\alpha}{r}e^{-\mu r}-\frac{\sigma}{2^{3/4}\Gamma(3/4)}\left(\frac{r}{\mu}\right)^{1/2}K_{1/4}[(\mu r)^2],
\end{eqnarray}
witr $\alpha=\frac{\pi}{12}$, $\sigma=0.2$\ GeV$^2$~\cite{Satz:2005hx}. The $\Gamma$\ and $K$ above are the gamma function and the modified Bessel function, respectively. We have dropped the constant term that does not vanish at infinity for simplicity in $V(r)$. The temperature $T$ dependence comes from the screening mass $\mu$. We fit the free energy of heavy quarks by the lattice QCD~\cite{Kaczmarek:2002mc, Satz:2005hx}, and parameterize the screening mass $\mu$ (scaled by $\sqrt{\sigma}$) as
\begin{eqnarray}
   \frac{\mu(\bar{T})}{\sqrt{\sigma}}&=& s\bar{T}+a\sigma_t\sqrt{\frac{\pi}{2}}\left[\erf\left(\frac{b}{\sqrt{2}\sigma_t}\right)-\erf\left(\frac{b-\bar{T}}{\sqrt{2}\sigma_t}\right)\right],
\end{eqnarray}
with $\bar{T}=T/T_c$, $s=0.587$, $a=2.150$, $b=1.054$,  $\sigma_t=0.07379$, and the error function $\erf(z)=\frac{2}{\sqrt{\pi}}\int_0^{z} e^{-x^2}\dd x$. Here $T_c$\ is the critical temperature of the phase transition.

\begin{figure}[!hbt]
    \centering
    \includegraphics[width=0.5\textwidth]{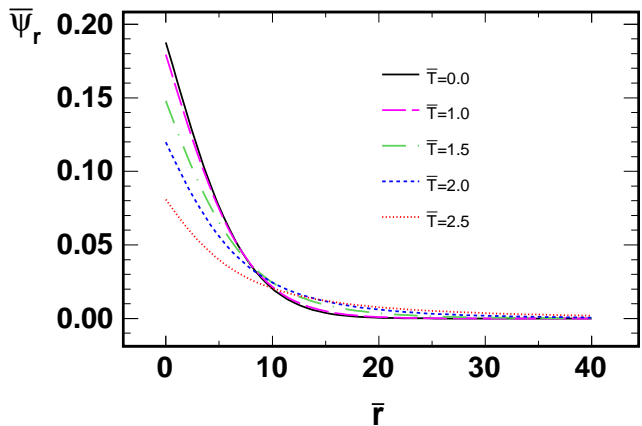}
    \caption{(Color Online)Scaled radial wave functions $\bar{\psi}_r=m_b^{-\frac{3}{2}}\psi$\ of $\Upsilon(1S)$\ as a function of scaled radius $\bar{r}=m_b r$, at different scaled temperature $\bar{T}=T/T_c$.}
    \label{fg_wf}
\end{figure}

The radial eigen wave function is shown in Fig.~\ref{fg_wf}. To be dimensionless, we scaled the radius and the wave function as $\bar{r}=m_b r$, and $\bar{\psi}_r=m_b^{-3/2}\psi_r$, respectively, resulting in $\int \left|\bar{\psi}_r\right|^2 \bar{r}^2\dd \bar{r}=1$. It can be seen that the wave function of $\Upsilon(1S)$\ at $T=T_c$\ is similar to that at $T=0$, while it becomes more and more broad at higher and higher temperature. The dissociation temperature is $T_d\approx3T_c$.

\begin{figure}[!hbt]
    \centering
    \includegraphics[width=0.5\textwidth]{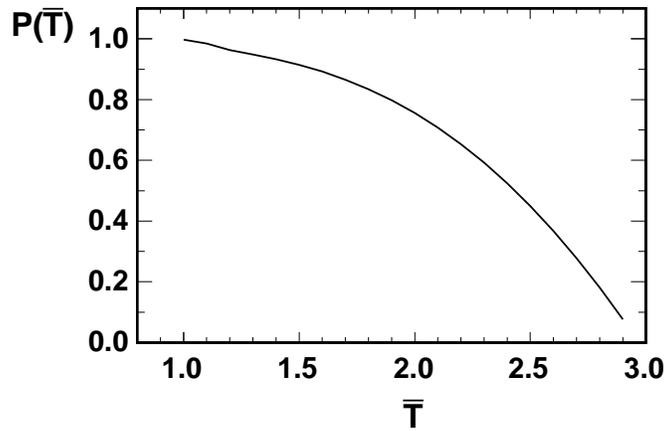}
    \caption{Transition probability $P$ [defined in Eq.~(\ref{eq_5})] of a $\Upsilon(1S)$\ from temperature $0$ to temperature $T$ as a function of scaled temperature $\bar{T}=T/T_c$.}
    \label{fg_prob}
\end{figure}

The transition probability from a $\Upsilon(1S)$ at zero temperature to that at $T$ is 
\begin{eqnarray}
   P(\bar{T})=\left|\left\langle \psi(T)|\psi(0)\right\rangle\right|^2,\label{eq_5}
\end{eqnarray}
which is shown as a function of $\bar{T}=T/T_c$ in Fig.~\ref{fg_prob}. It decreases with $\bar{T}$ monototically, since the overlap between the wave function at finite temperature and that at zero temperature becomes small when $\bar{T}$ increases.  It is very close to unit at $\bar{T}=1$ as already indicated by Fig.~\ref{fg_wf}, and it vanishes at $T_d\approx3T_c$.

\begin{figure}[!hbt]
\centering
\includegraphics[width=0.45\textwidth]{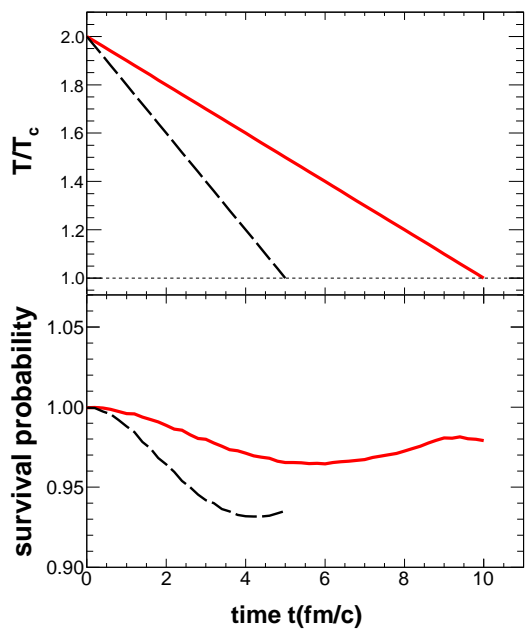}
\caption{(Color Online) 
Upper panel: Different cooling systems with the medium temperature decreasing linearly with time. 
Lower panel: 
Time evolution of survival probability of $\Upsilon(1S)$\ at different cooling speed calculated by Schroedinger equation with an initial $\Upsilon(1S)$ at its eigen state at the initial temperature.
}
\hspace{-0.1mm}
\label{fig-decrease}
\label{fg_sur_linear}
\end{figure}
Now we check the adiabatic approximation. At RHIC energy, the highest temperature of the fireball is  around $2T_c$ when the system reaches local thermal equilibrium. We suppose that the temperature decreases with time linearly from $2T_c$ to $T_c$, and evolve the wave function $\psi({\bm r},t)$\ of a $\Upsilon(1S)$ by Eq.~(\ref{eq_1}) with its initial condition as an eigen $\Upsilon(1S)$\ at $2T_c$. The survival probability as a function of time is shown in Fig.~\ref{fg_sur_linear}. The typical time for the fireball to cool down to $T_c$\ is $5\sim10$\ fm/$c$. As one can see from the figure, the survival probability is about $0.98$ when the evolution time is $10$\ fm/$c$, which implies that the adiabatic approximation is very good in such a case. Even if we take a lower value of $5$\ fm/$c$, the survival probability $0.93$\ is obviously larger than $P(2.0)=0.76$\ shown in Fig.~\ref{fg_prob}. This result is qualitatively consistent with the result in Ref.~\cite{Boyd:2019arx}, where the adiabatic approximation is examinded for $\Upsilon(1S)$\ at LHC energy with a finite dissociation rate.

\begin{figure}[!hbt]
    \centering
    \includegraphics[width=0.5\textwidth]{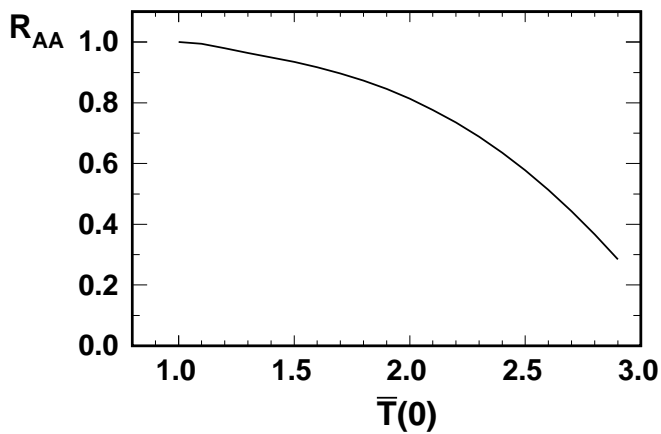}
    \caption{The nuclear modification factor $R_{AA}$\ in central Au+Au collisions due to the heating dissociation effect as a function of the scaled temperature $\bar{T}({\bm 0})$ at the central of the fireball.}
    \label{fg_raa}
\end{figure}

Now we include the spacial distribution of temperature. In practice the temperature is not uniform in space. The temperature is high in the center of the fireball, while it is low in peripheral regions. Therefore the survival probability for $\Upsilon(1S)$\ is an average of all produced $\Upsilon(1S)$s. Since the production of $\Upsilon(1S)$\ is a hard process, we assume that the density of produced $\Upsilon(1S)$\ is proportional to the number density of binary collisions $n_c({\bm x}_T)$ at transverse coordinate ${\bm x}_T$. Therefore we have
\begin{eqnarray}
   R_{AA}&=& \frac{\int P(\bar{T}({\bm x}_T))\dd N_{\Upsilon(1S)}}{\int \dd N_{\Upsilon(1S)}}=\frac{\int P(\bar{T}({\bm x}_T))n_c({\bm x}_T)\dd {\bm x}_T}{\int n_c({\bm x}_T)\dd {\bm x}_T}.\label{eq_6}
\end{eqnarray}
We assume that the entropy density $s$ is proportional to the density of the number of participants $n_p$, and regard the hot medium as ideal gas, so that the entropy density is also proportional to $T^3$. As a result, the spacial distribution of temperature is
\begin{eqnarray}
   \bar{T}({\bm x}_T)&=& \bar{T}({\bm 0})\left(\frac{n_p({\bm x}_T)}{n_p({\bm 0})}\right)^{1/3},\label{eq_7}
\end{eqnarray}
where $\bar{T}({\bm x}_T)$\ is the scaled local temperature $T/T_c$\ at ${\bm x}_T$, and $\bar{T}({\bm 0})$\ is the scaled local temperature at ${\bm x}_T={\bm 0}$. In central collisions, the number density of participants $n_p$\ and number density of binary collisions $n_c$ are
\begin{eqnarray}
   n_p({\bm x}_T)&=& 2\mathcal{T}(x_T)\left[1-e^{-\sigma_{NN}\mathcal{T}(x_T)}\right],\\
   n_c({\bm x}_T)&=& \sigma_{NN}\mathcal{T}^2(x_T),
\end{eqnarray}
where $\sigma_{NN}$\ is the inelastic cross section of nucleons, and $\mathcal{T}(x_T)$\ is the thickness function of a gold nucleus. For simplicity, we take a sharp-cut-off thickness function 
\begin{eqnarray}
   \mathcal{T}(x_T)&=& \frac{3A\sqrt{R^2-x_T^2}}{2\pi R^3},\label{eq_10}
\end{eqnarray}
where $R$\ and $A$\ are the radius and mass number of the nucleus, respectively. Substitute Eq.~(\ref{eq_7}-\ref{eq_10}) to Eq.~(\ref{eq_6}), we obtain the nuclear modification factor in central collisions
\begin{eqnarray}
   R_{AA}=4\int_0^1P\left(\bar{T}(0)\sqrt[\uproot{3}3]{x\frac{1-e^{-N_mx}}{1-e^{-N_m}}}\right)x^3\dd x,\label{eq_11}
\end{eqnarray}
with $N_m=\sigma_{NN}\mathcal{T}(x_T=0)=3\sigma_{NN}A/(2\pi R^2)$, and
\begin{eqnarray}
   P(\bar{T})&=& \left\{\begin{array}{ll}0,&\bar{T}>T_d/T_c,\\\left|\langle \psi(T)|\psi(0)\rangle\right|^2,&1<\bar{T}<T_d/T_c,\\1,&\bar{T}<1.\end{array}\right.
\end{eqnarray}
where we have taken $P=1$\ below $T_c$ as an approximation. We take  $R=6.38$ fm and $A=197$ for gold~\cite{nuclgeo}, and $\sigma_{NN}=41$ mb at RHIC energy~\cite{Zhu:2004nw}. The $R_{AA}$\ as a function of $\bar{T}({\bm 0})$\ is shown in Fig.~\ref{fg_raa}. It can be seen that the $R_{AA}$\ is above $0.9$\ if the central temperature $T(0)$\ is lower than $1.6T_c$, while it is below $0.8$ when $T(0)$\ is higher than $2.1T_c$. One can expect that this effect is not negligible at the RHIC and is remarkable at the LHC. The factor $x^3$\ in Eq.~(\ref{eq_11}) comes from two facts: 1) more $\Upsilon(1S)$s are produced in central of the fireball, and 2) the thickness changes slow with radius in central of the fireball. As a result, the $R_{AA}$\ relies more on the survival probability $P$ in the center of the fireball, that is at $x_T=0$. Therefore the qualitative behavior of the $R_{AA}$\ in Fig.~\ref{fg_raa} is similar to the $P$\ in Fig.~\ref{fg_prob}, and they are quantitatively similar when $P(\bar{T}({\bm 0}))$\ is large.

We have two remarks on this result. 1) Even if the width (or dissociation rate) $\Gamma$\ vanishes at finite temperature, there is a fast heating dissociation effect for $\Upsilon(1S)$ suppression, which is not carefully considered before. 2) If the width of $\Upsilon(1S)$\ is negligible as in some calculations, then the heating dissociation of $\Upsilon(1S)$\ can be used as a thermometer to detect the temperature of the fireball at early time, and it is not sensitive to the temperature later on. It is necessary to clarify that such a temperature measured by $\Upsilon(1S)$\ should never be interpreted as the highest temperature of the fireball, but the temperature a $\Upsilon(1S)$ feels. As a matter of fact, the highest temperature at very early time is not well defined and the change of the temperature at very early time is so quick that the adiabatic theorem breaks, which means the $\Upsilon(1S)$\ may not feel the temperature before the temperature drops down relatively slowly. Actually the most interesting temperature is not the high and short-lived temperature at the very beginning, but the temperature that can be felt by particles. In this sense, the $\Upsilon(1S)$-felt temperature  of the medium is more meaningful.

In summary we discussed a new mechanism on $\Upsilon(1S)$ dissociation which is due to the fast heating process at the early stage of the fireball instead of a non-zero width in a steady hot medium. Because of such a fast heating dissociaiton, the suppression of $\Upsilon(1S)$ is observable at RHIC energy even if the width of $\Upsilon(1S)$\ at finite temperature is zero,  and such a mechanism may be used as a measure of the temperature of the fireball at early time.

Acknowlegements: the work is supported by the NSFC under the Grant No.s 11547043, 11705125 and by the ``Qinggu'' project of Tianjin University.
\bibliography{refs}

\begin{thebibliography}{37}
\expandafter\ifx\csname natexlab\endcsname\relax\def\natexlab#1{#1}\fi
\expandafter\ifx\csname bibnamefont\endcsname\relax
  \def\bibnamefont#1{#1}\fi
\expandafter\ifx\csname bibfnamefont\endcsname\relax
  \def\bibfnamefont#1{#1}\fi
\expandafter\ifx\csname citenamefont\endcsname\relax
  \def\citenamefont#1{#1}\fi
\expandafter\ifx\csname url\endcsname\relax
  \def\url#1{\texttt{#1}}\fi
\expandafter\ifx\csname urlprefix\endcsname\relax\def\urlprefix{URL }\fi
\providecommand{\bibinfo}[2]{#2}
\providecommand{\eprint}[2][]{\url{#2}}

\bibitem[{\citenamefont{Hagedorn}(1965)}]{Hagedorn:346206}
\bibinfo{author}{\bibfnamefont{R.}~\bibnamefont{Hagedorn}},
  \bibinfo{journal}{Nuovo Cimento, Suppl.} \textbf{\bibinfo{volume}{3}},
  \bibinfo{pages}{147} (\bibinfo{year}{1965}),
  \urlprefix\url{http://cds.cern.ch/record/346206}.

\bibitem[{\citenamefont{Karsch et~al.}(2001)\citenamefont{Karsch, Laermann, and
  Peikert}}]{Karsch:2000kv}
\bibinfo{author}{\bibfnamefont{F.}~\bibnamefont{Karsch}},
  \bibinfo{author}{\bibfnamefont{E.}~\bibnamefont{Laermann}}, \bibnamefont{and}
  \bibinfo{author}{\bibfnamefont{A.}~\bibnamefont{Peikert}},
  \bibinfo{journal}{Nucl. Phys.} \textbf{\bibinfo{volume}{B605}},
  \bibinfo{pages}{579} (\bibinfo{year}{2001}), \eprint{hep-lat/0012023}.

\bibitem[{\citenamefont{Ding et~al.}(2015)\citenamefont{Ding, Karsch, and
  Mukherjee}}]{Ding:2015ona}
\bibinfo{author}{\bibfnamefont{H.-T.} \bibnamefont{Ding}},
  \bibinfo{author}{\bibfnamefont{F.}~\bibnamefont{Karsch}}, \bibnamefont{and}
  \bibinfo{author}{\bibfnamefont{S.}~\bibnamefont{Mukherjee}},
  \bibinfo{journal}{Int. J. Mod. Phys.} \textbf{\bibinfo{volume}{E24}},
  \bibinfo{pages}{1530007} (\bibinfo{year}{2015}), \eprint{1504.05274}.

\bibitem[{\citenamefont{Pisarski and Wilczek}(1984)}]{Pisarski:1983ms}
\bibinfo{author}{\bibfnamefont{R.~D.} \bibnamefont{Pisarski}} \bibnamefont{and}
  \bibinfo{author}{\bibfnamefont{F.}~\bibnamefont{Wilczek}},
  \bibinfo{journal}{Phys. Rev.} \textbf{\bibinfo{volume}{D29}},
  \bibinfo{pages}{338} (\bibinfo{year}{1984}).

\bibitem[{\citenamefont{Fukushima and Hatsuda}(2011)}]{Fukushima:2010bq}
\bibinfo{author}{\bibfnamefont{K.}~\bibnamefont{Fukushima}} \bibnamefont{and}
  \bibinfo{author}{\bibfnamefont{T.}~\bibnamefont{Hatsuda}},
  \bibinfo{journal}{Rept. Prog. Phys.} \textbf{\bibinfo{volume}{74}},
  \bibinfo{pages}{014001} (\bibinfo{year}{2011}), \eprint{1005.4814}.

\bibitem[{\citenamefont{Stephanov et~al.}(1998)\citenamefont{Stephanov,
  Rajagopal, and Shuryak}}]{Stephanov:1998dy}
\bibinfo{author}{\bibfnamefont{M.~A.} \bibnamefont{Stephanov}},
  \bibinfo{author}{\bibfnamefont{K.}~\bibnamefont{Rajagopal}},
  \bibnamefont{and} \bibinfo{author}{\bibfnamefont{E.~V.}
  \bibnamefont{Shuryak}}, \bibinfo{journal}{Phys. Rev. Lett.}
  \textbf{\bibinfo{volume}{81}}, \bibinfo{pages}{4816} (\bibinfo{year}{1998}),
  \eprint{hep-ph/9806219}.

\bibitem[{\citenamefont{Kharzeev and Zhitnitsky}(2007)}]{Kharzeev:2007tn}
\bibinfo{author}{\bibfnamefont{D.}~\bibnamefont{Kharzeev}} \bibnamefont{and}
  \bibinfo{author}{\bibfnamefont{A.}~\bibnamefont{Zhitnitsky}},
  \bibinfo{journal}{Nucl. Phys.} \textbf{\bibinfo{volume}{A797}},
  \bibinfo{pages}{67} (\bibinfo{year}{2007}), \eprint{0706.1026}.

\bibitem[{\citenamefont{Gonin et~al.}(1996)}]{Gonin:1996wn}
\bibinfo{author}{\bibfnamefont{M.}~\bibnamefont{Gonin}} \bibnamefont{et~al.}
  (\bibinfo{collaboration}{NA50}), \bibinfo{journal}{Nucl. Phys.}
  \textbf{\bibinfo{volume}{A610}}, \bibinfo{pages}{404c}
  (\bibinfo{year}{1996}).

\bibitem[{\citenamefont{Luo and Xu}(2017)}]{Luo:2017faz}
\bibinfo{author}{\bibfnamefont{X.}~\bibnamefont{Luo}} \bibnamefont{and}
  \bibinfo{author}{\bibfnamefont{N.}~\bibnamefont{Xu}}, \bibinfo{journal}{Nucl.
  Sci. Tech.} \textbf{\bibinfo{volume}{28}}, \bibinfo{pages}{112}
  (\bibinfo{year}{2017}), \eprint{1701.02105}.

\bibitem[{\citenamefont{Adcox et~al.}(2002)}]{Adcox:2001jp}
\bibinfo{author}{\bibfnamefont{K.}~\bibnamefont{Adcox}} \bibnamefont{et~al.}
  (\bibinfo{collaboration}{PHENIX}), \bibinfo{journal}{Phys. Rev. Lett.}
  \textbf{\bibinfo{volume}{88}}, \bibinfo{pages}{022301}
  (\bibinfo{year}{2002}), \eprint{nucl-ex/0109003}.

\bibitem[{\citenamefont{Adare et~al.}(2007)}]{Adare:2006ti}
\bibinfo{author}{\bibfnamefont{A.}~\bibnamefont{Adare}} \bibnamefont{et~al.}
  (\bibinfo{collaboration}{PHENIX}), \bibinfo{journal}{Phys. Rev. Lett.}
  \textbf{\bibinfo{volume}{98}}, \bibinfo{pages}{162301}
  (\bibinfo{year}{2007}), \eprint{nucl-ex/0608033}.

\bibitem[{\citenamefont{Chatrchyan et~al.}(2011)}]{Chatrchyan:2011sx}
\bibinfo{author}{\bibfnamefont{S.}~\bibnamefont{Chatrchyan}}
  \bibnamefont{et~al.} (\bibinfo{collaboration}{CMS}), \bibinfo{journal}{Phys.
  Rev.} \textbf{\bibinfo{volume}{C84}}, \bibinfo{pages}{024906}
  (\bibinfo{year}{2011}), \eprint{1102.1957}.

\bibitem[{\citenamefont{Andronic et~al.}(2009)\citenamefont{Andronic,
  Braun-Munzinger, and Stachel}}]{Andronic:2009qf}
\bibinfo{author}{\bibfnamefont{A.}~\bibnamefont{Andronic}},
  \bibinfo{author}{\bibfnamefont{P.}~\bibnamefont{Braun-Munzinger}},
  \bibnamefont{and} \bibinfo{author}{\bibfnamefont{J.}~\bibnamefont{Stachel}},
  \bibinfo{journal}{Acta Phys. Polon.} \textbf{\bibinfo{volume}{B40}},
  \bibinfo{pages}{1005} (\bibinfo{year}{2009}), \eprint{0901.2909}.

\bibitem[{\citenamefont{Bellwied et~al.}(2019)\citenamefont{Bellwied,
  Noronha-Hostler, Parotto, Vazquez, Ratti, and
  Stafford}}]{bellwied2019extracting}
\bibinfo{author}{\bibfnamefont{R.}~\bibnamefont{Bellwied}},
  \bibinfo{author}{\bibfnamefont{J.}~\bibnamefont{Noronha-Hostler}},
  \bibinfo{author}{\bibfnamefont{P.}~\bibnamefont{Parotto}},
  \bibinfo{author}{\bibfnamefont{I.~P.} \bibnamefont{Vazquez}},
  \bibinfo{author}{\bibfnamefont{C.}~\bibnamefont{Ratti}}, \bibnamefont{and}
  \bibinfo{author}{\bibfnamefont{J.}~\bibnamefont{Stafford}},
  \emph{\bibinfo{title}{Extracting the strangeness freeze-out temperature from
  net-kaon data at rhic}} (\bibinfo{year}{2019}), \eprint{1904.12711}.

\bibitem[{\citenamefont{Matsui and Satz}(1986)}]{Matsui:1986dk}
\bibinfo{author}{\bibfnamefont{T.}~\bibnamefont{Matsui}} \bibnamefont{and}
  \bibinfo{author}{\bibfnamefont{H.}~\bibnamefont{Satz}},
  \bibinfo{journal}{Phys. Lett.} \textbf{\bibinfo{volume}{B178}},
  \bibinfo{pages}{416} (\bibinfo{year}{1986}).

\bibitem[{\citenamefont{Karsch et~al.}(2006)\citenamefont{Karsch, Kharzeev, and
  Satz}}]{Karsch:2005nk}
\bibinfo{author}{\bibfnamefont{F.}~\bibnamefont{Karsch}},
  \bibinfo{author}{\bibfnamefont{D.}~\bibnamefont{Kharzeev}}, \bibnamefont{and}
  \bibinfo{author}{\bibfnamefont{H.}~\bibnamefont{Satz}},
  \bibinfo{journal}{Phys. Lett.} \textbf{\bibinfo{volume}{B637}},
  \bibinfo{pages}{75} (\bibinfo{year}{2006}), \eprint{hep-ph/0512239}.

\bibitem[{\citenamefont{Peskin}(1979)}]{Peskin:1979va}
\bibinfo{author}{\bibfnamefont{M.~E.} \bibnamefont{Peskin}},
  \bibinfo{journal}{Nucl. Phys.} \textbf{\bibinfo{volume}{B156}},
  \bibinfo{pages}{365} (\bibinfo{year}{1979}).

\bibitem[{\citenamefont{Xu et~al.}(1996)\citenamefont{Xu, Kharzeev, Satz, and
  Wang}}]{Xu:1995eb}
\bibinfo{author}{\bibfnamefont{X.-M.} \bibnamefont{Xu}},
  \bibinfo{author}{\bibfnamefont{D.}~\bibnamefont{Kharzeev}},
  \bibinfo{author}{\bibfnamefont{H.}~\bibnamefont{Satz}}, \bibnamefont{and}
  \bibinfo{author}{\bibfnamefont{X.-N.} \bibnamefont{Wang}},
  \bibinfo{journal}{Phys. Rev.} \textbf{\bibinfo{volume}{C53}},
  \bibinfo{pages}{3051} (\bibinfo{year}{1996}), \eprint{hep-ph/9511331}.

\bibitem[{\citenamefont{Chen and He}(2018)}]{Chen:2018dqg}
\bibinfo{author}{\bibfnamefont{S.}~\bibnamefont{Chen}} \bibnamefont{and}
  \bibinfo{author}{\bibfnamefont{M.}~\bibnamefont{He}}, \bibinfo{journal}{Phys.
  Lett.} \textbf{\bibinfo{volume}{B786}}, \bibinfo{pages}{260}
  (\bibinfo{year}{2018}), \eprint{1805.06346}.

\bibitem[{\citenamefont{Yao and Muller}(2019)}]{Yao:2018sgn}
\bibinfo{author}{\bibfnamefont{X.}~\bibnamefont{Yao}} \bibnamefont{and}
  \bibinfo{author}{\bibfnamefont{B.}~\bibnamefont{Muller}},
  \bibinfo{journal}{Phys. Rev.} \textbf{\bibinfo{volume}{D100}},
  \bibinfo{pages}{014008} (\bibinfo{year}{2019}), \eprint{1811.09644}.

\bibitem[{\citenamefont{Grandchamp and Rapp}(2001)}]{Grandchamp:2001pf}
\bibinfo{author}{\bibfnamefont{L.}~\bibnamefont{Grandchamp}} \bibnamefont{and}
  \bibinfo{author}{\bibfnamefont{R.}~\bibnamefont{Rapp}},
  \bibinfo{journal}{Phys. Lett.} \textbf{\bibinfo{volume}{B523}},
  \bibinfo{pages}{60} (\bibinfo{year}{2001}), \eprint{hep-ph/0103124}.

\bibitem[{\citenamefont{Thews et~al.}(2001)\citenamefont{Thews, Schroedter, and
  Rafelski}}]{Thews:2000rj}
\bibinfo{author}{\bibfnamefont{R.~L.} \bibnamefont{Thews}},
  \bibinfo{author}{\bibfnamefont{M.}~\bibnamefont{Schroedter}},
  \bibnamefont{and} \bibinfo{author}{\bibfnamefont{J.}~\bibnamefont{Rafelski}},
  \bibinfo{journal}{Phys. Rev.} \textbf{\bibinfo{volume}{C63}},
  \bibinfo{pages}{054905} (\bibinfo{year}{2001}), \eprint{hep-ph/0007323}.

\bibitem[{\citenamefont{Yan et~al.}(2006)\citenamefont{Yan, Zhuang, and
  Xu}}]{Yan:2006ve}
\bibinfo{author}{\bibfnamefont{L.}~\bibnamefont{Yan}},
  \bibinfo{author}{\bibfnamefont{P.}~\bibnamefont{Zhuang}}, \bibnamefont{and}
  \bibinfo{author}{\bibfnamefont{N.}~\bibnamefont{Xu}}, \bibinfo{journal}{Phys.
  Rev. Lett.} \textbf{\bibinfo{volume}{97}}, \bibinfo{pages}{232301}
  (\bibinfo{year}{2006}), \eprint{nucl-th/0608010}.

\bibitem[{\citenamefont{Yao and Mehen}(2019)}]{Yao:2018nmy}
\bibinfo{author}{\bibfnamefont{X.}~\bibnamefont{Yao}} \bibnamefont{and}
  \bibinfo{author}{\bibfnamefont{T.}~\bibnamefont{Mehen}},
  \bibinfo{journal}{Phys. Rev.} \textbf{\bibinfo{volume}{D99}},
  \bibinfo{pages}{096028} (\bibinfo{year}{2019}), \eprint{1811.07027}.

\bibitem[{\citenamefont{Zhu et~al.}(2005)\citenamefont{Zhu, Zhuang, and
  Xu}}]{Zhu:2004nw}
\bibinfo{author}{\bibfnamefont{X.}~\bibnamefont{Zhu}},
  \bibinfo{author}{\bibfnamefont{P.}~\bibnamefont{Zhuang}}, \bibnamefont{and}
  \bibinfo{author}{\bibfnamefont{N.}~\bibnamefont{Xu}}, \bibinfo{journal}{Phys.
  Lett.} \textbf{\bibinfo{volume}{B607}}, \bibinfo{pages}{107}
  (\bibinfo{year}{2005}), \eprint{nucl-th/0411093}.

\bibitem[{\citenamefont{Song et~al.}(2008)\citenamefont{Song, Park, Lee, and
  Wong}}]{Song:2007gm}
\bibinfo{author}{\bibfnamefont{T.}~\bibnamefont{Song}},
  \bibinfo{author}{\bibfnamefont{Y.}~\bibnamefont{Park}},
  \bibinfo{author}{\bibfnamefont{S.~H.} \bibnamefont{Lee}}, \bibnamefont{and}
  \bibinfo{author}{\bibfnamefont{C.-Y.} \bibnamefont{Wong}},
  \bibinfo{journal}{Phys. Lett.} \textbf{\bibinfo{volume}{B659}},
  \bibinfo{pages}{621} (\bibinfo{year}{2008}), \eprint{0709.0794}.

\bibitem[{\citenamefont{Zhao and Rapp}(2011)}]{Zhao:2011cv}
\bibinfo{author}{\bibfnamefont{X.}~\bibnamefont{Zhao}} \bibnamefont{and}
  \bibinfo{author}{\bibfnamefont{R.}~\bibnamefont{Rapp}},
  \bibinfo{journal}{Nucl. Phys.} \textbf{\bibinfo{volume}{A859}},
  \bibinfo{pages}{114} (\bibinfo{year}{2011}), \eprint{1102.2194}.

\bibitem[{\citenamefont{Brambilla et~al.}(2000)\citenamefont{Brambilla, Pineda,
  Soto, and Vairo}}]{Brambilla:1999xf}
\bibinfo{author}{\bibfnamefont{N.}~\bibnamefont{Brambilla}},
  \bibinfo{author}{\bibfnamefont{A.}~\bibnamefont{Pineda}},
  \bibinfo{author}{\bibfnamefont{J.}~\bibnamefont{Soto}}, \bibnamefont{and}
  \bibinfo{author}{\bibfnamefont{A.}~\bibnamefont{Vairo}},
  \bibinfo{journal}{Nucl. Phys.} \textbf{\bibinfo{volume}{B566}},
  \bibinfo{pages}{275} (\bibinfo{year}{2000}), \eprint{hep-ph/9907240}.

\bibitem[{\citenamefont{Escobedo et~al.}(2013)\citenamefont{Escobedo,
  Giannuzzi, Mannarelli, and Soto}}]{Escobedo:2013tca}
\bibinfo{author}{\bibfnamefont{M.~A.} \bibnamefont{Escobedo}},
  \bibinfo{author}{\bibfnamefont{F.}~\bibnamefont{Giannuzzi}},
  \bibinfo{author}{\bibfnamefont{M.}~\bibnamefont{Mannarelli}},
  \bibnamefont{and} \bibinfo{author}{\bibfnamefont{J.}~\bibnamefont{Soto}},
  \bibinfo{journal}{Phys. Rev.} \textbf{\bibinfo{volume}{D87}},
  \bibinfo{pages}{114005} (\bibinfo{year}{2013}), \eprint{1304.4087}.

\bibitem[{\citenamefont{Mishra et~al.}(2019)\citenamefont{Mishra, Jahan~CS,
  Kesarwani, Raval, Kumar, and Meena}}]{CS:2018mag}
\bibinfo{author}{\bibfnamefont{A.}~\bibnamefont{Mishra}},
  \bibinfo{author}{\bibfnamefont{A.}~\bibnamefont{Jahan~CS}},
  \bibinfo{author}{\bibfnamefont{S.}~\bibnamefont{Kesarwani}},
  \bibinfo{author}{\bibfnamefont{H.}~\bibnamefont{Raval}},
  \bibinfo{author}{\bibfnamefont{S.}~\bibnamefont{Kumar}}, \bibnamefont{and}
  \bibinfo{author}{\bibfnamefont{J.}~\bibnamefont{Meena}},
  \bibinfo{journal}{Eur. Phys. J.} \textbf{\bibinfo{volume}{A55}},
  \bibinfo{pages}{99} (\bibinfo{year}{2019}), \eprint{1812.07397}.

\bibitem[{\citenamefont{Liu}(2009)}]{Liu:2009wa}
\bibinfo{author}{\bibfnamefont{H.}~\bibnamefont{Liu}}
  (\bibinfo{collaboration}{STAR}), \bibinfo{journal}{Nucl. Phys.}
  \textbf{\bibinfo{volume}{A830}}, \bibinfo{pages}{235c}
  (\bibinfo{year}{2009}), \eprint{0907.4538}.

\bibitem[{\citenamefont{Du et~al.}(2017)\citenamefont{Du, He, and
  Rapp}}]{Du_2017}
\bibinfo{author}{\bibfnamefont{X.}~\bibnamefont{Du}},
  \bibinfo{author}{\bibfnamefont{M.}~\bibnamefont{He}}, \bibnamefont{and}
  \bibinfo{author}{\bibfnamefont{R.}~\bibnamefont{Rapp}},
  \bibinfo{journal}{Physical Review} \textbf{\bibinfo{volume}{C96}}
  (\bibinfo{year}{2017}), ISSN \bibinfo{issn}{2469-9993},
  \urlprefix\url{http://dx.doi.org/10.1103/PhysRevC.96.054901}.

\bibitem[{\citenamefont{Katz and Gossiaux}(2016)}]{Katz:2015qja}
\bibinfo{author}{\bibfnamefont{R.}~\bibnamefont{Katz}} \bibnamefont{and}
  \bibinfo{author}{\bibfnamefont{P.~B.} \bibnamefont{Gossiaux}},
  \bibinfo{journal}{Annals Phys.} \textbf{\bibinfo{volume}{368}},
  \bibinfo{pages}{267} (\bibinfo{year}{2016}), \eprint{1504.08087}.

\bibitem[{\citenamefont{Satz}(2006)}]{Satz:2005hx}
\bibinfo{author}{\bibfnamefont{H.}~\bibnamefont{Satz}}, \bibinfo{journal}{J.
  Phys.} \textbf{\bibinfo{volume}{G32}}, \bibinfo{pages}{R25}
  (\bibinfo{year}{2006}), \eprint{hep-ph/0512217}.

\bibitem[{\citenamefont{Kaczmarek et~al.}(2002)\citenamefont{Kaczmarek, Karsch,
  Petreczky, and Zantow}}]{Kaczmarek:2002mc}
\bibinfo{author}{\bibfnamefont{O.}~\bibnamefont{Kaczmarek}},
  \bibinfo{author}{\bibfnamefont{F.}~\bibnamefont{Karsch}},
  \bibinfo{author}{\bibfnamefont{P.}~\bibnamefont{Petreczky}},
  \bibnamefont{and} \bibinfo{author}{\bibfnamefont{F.}~\bibnamefont{Zantow}},
  \bibinfo{journal}{Phys. Lett.} \textbf{\bibinfo{volume}{B543}},
  \bibinfo{pages}{41} (\bibinfo{year}{2002}), \eprint{hep-lat/0207002}.

\bibitem[{\citenamefont{Boyd et~al.}(2019)\citenamefont{Boyd, Cook, Islam, and
  Strickland}}]{Boyd:2019arx}
\bibinfo{author}{\bibfnamefont{J.}~\bibnamefont{Boyd}},
  \bibinfo{author}{\bibfnamefont{T.}~\bibnamefont{Cook}},
  \bibinfo{author}{\bibfnamefont{A.}~\bibnamefont{Islam}}, \bibnamefont{and}
  \bibinfo{author}{\bibfnamefont{M.}~\bibnamefont{Strickland}},
  \bibinfo{journal}{Phys. Rev.} \textbf{\bibinfo{volume}{D100}},
  \bibinfo{pages}{076019} (\bibinfo{year}{2019}), \eprint{1905.05676}.

\bibitem[{\citenamefont{De~Vries et~al.}(1987)\citenamefont{De~Vries, De~Jager,
  and De~Vries}}]{nuclgeo}
\bibinfo{author}{\bibfnamefont{H.}~\bibnamefont{De~Vries}},
  \bibinfo{author}{\bibfnamefont{C.}~\bibnamefont{De~Jager}}, \bibnamefont{and}
  \bibinfo{author}{\bibfnamefont{C.}~\bibnamefont{De~Vries}},
  \bibinfo{journal}{Atom. Data Nucl. Data Tabl.} \textbf{\bibinfo{volume}{36}},
  \bibinfo{pages}{495} (\bibinfo{year}{1987}).

\end{thebibliography}
\end{document}